\documentclass[10pt,a4paper]{article}
\usepackage{eurosym}
\usepackage{amsmath,amssymb,amsfonts}
\usepackage{graphicx}
\usepackage{epstopdf}
\usepackage{hyperref}
\usepackage{cite}
\usepackage{bm}
\usepackage{setspace}
\usepackage{appendix}

\begin{document}

\title{Transients in porous media: exact and modelled time-domain Green's
functions}
\author{J. Kergomard\thanks{%
Tel 33 491164381, Fax 33 491228248, kergomard@lma.cnrs-mrs.fr} (a), D.
Lafarge (b), J. Gilbert (b) \\
%EndAName
\\
(a) LMA, CNRS, UPR 7051, Aix-Marseille Univ, Centrale Marseille, \\
F-13402 Marseille Cedex 20, France\\
(b) LUNAM Universit\'{e}, CNRS, UMR 6613, \\
Laboratoire d'Acoustique de l'Universit\'{e} du Maine, Avenue Olivier
Messiaen,\\
72085 Le Mans Cedex 9, France.\\
}
\maketitle

\begin{abstract}
Time domain responses of porous media have been studied by some authors, but
generally the possible descriptions are given in the frequency domain. The
aim of this paper, limited to materials with rigid skeleton considered as
equivalent fluids, is to compare in time domain different descriptions by
Johnson-Allard ($JA$) as well as by Pride-Lafarge ($PL$), with: $i$) some
analytical approximate formulas based upon asymptotic high-frequency
expansion ; $ii$) the exact formula by Zwikker and Kosten for the case of
cylindrical pores. The paper starts with a construction analysis of the
models ($JA$ and $PL$). Then, the Green's function in the time domain is
defined, written in scaled form, and shown to exhibit interesting properties
of materials. In particular, a so-far overlooked decay length $\mathcal{L}$
describing a high-frequency attenuation-without-distortion effect is
identified in terms of Brown's tortuosity, Johnson's and Allard's known
characteristic viscous and thermal lengths, and two, unknown in general,
characteristic viscous and thermal surfaces. The numerical computation of
the Green's function is done by FFT, with some precautions, because of the
importance of the higher frequencies on the response shape: the substraction
of the diffusive (low frequencies) approximation largely improves the
results of the FFT. The $PL$ description is shown to be the best
full-frequency general model with remaining small discrepancies due to
unsatisfactory account of the mentioned surface-parameters.
%Finally, a possible relevance of our results for the acoustic characterization of materials is briefly discussed.
\
\end{abstract}

%(c)Laboratoire de Physique Th\'{e}orique, Facult\'{e} de Physique,\\
%USTHB, BP 32 El Alia, Bab Ezzouar, 16111, Algeria.}

%\listoffigures
\doublespacing

\singlespacing

\section{\protect\bigskip Introduction}

Linear wave propagation in homogeneous porous media saturated with a
viscothermal fluid such as ambient air, has been the subject of extensive
research. Traditionally, for the case when the wavelength is large compared
to the pore size, it has been described on the basis of the so-called
two-scale asymptotic homogenization theory \cite{bens}--\cite{burr}. For
materials with skeleton sufficiently heavy and/or rigid to be motionless,
this leads to an effective medium theory in which the medium permittivities
are two linear operators, one representing an effective density and the
other an effective compressibility \cite{norr}, \cite{lafarge}. These
operators, while nonlocal in time as a result of delayed responses due to
the losses (temporal dispersion), are given as local in space by the
homogenization process at the dominant order. In other words, no spatial
dispersion arises, so that the response of the material at a given
macroscopic point (physically, an elementary coarse-graining volume) is
determined by the history of the pertinent field variables at the given
point but not the neighboring points. As such, this description is a special
case not directly applicable to all geometries. It does not describe the
situations where structures in the form of Helmholtz resonators are present%
\footnote{\label{1f}In this case, if the microgeometry splits in parts with
pore sizes sufficiently different to imply different rescalings of the
microscopic governing equations in the different parts, a solution can still
be written using the principle of asymptotic homogenization \cite{boutin}.
Independently, a general nonlocal theory of propagation along a symmetry
axis in macroscopically homogeneous materials having arbitrary
microgeometry, has been recently proposed by one of the authors \cite{laf1}.
Finally, different high-frequencies extensions of the idea of two-scale
homogenization have also been introduced recently \cite{kap1}--\cite{bou1}.
All of these extensions lie outside the scope of the present paper, as
defined next.}. Nevertheless, most of the materials used in noise control
for sound absorption, do not present in their inner structure very different
pores sizes. In this paper we explicitly assume the absence of very
different pores sizes\footnote{%
This excludes resonators, as in a resonator the neck and cavity dimensions
differ by one order of magnitude at least.}, automatically ensuring at long
wavelengths the validity of the usual two-scale asymptotic homogenization.

It gives us explicit recipes to compute from microgeometry the effective
density and compressibility, complex functions of frequency but not of
wavenumber, in the above framework. In practice, the computation is not
possible in full detail; it is however not required to be done to arrive at
a relatively precise description. In absence of a complete knowledge of the
microgeometry, a widely used semi-phenomenological model which depends on a
small set of independently measurable geometrical parameters of the
structure, is given by the well-known formula of Johnson $et$ $al.$ \cite%
{joh} for the density, and likewise, the Champoux-Allard \cite{champ} or the
Lafarge $et$ $al.$ \cite{lafarge} formula for the compressibility\footnote{%
This last formula relies on a description of elastic-thermal effects
entirely analogous to Johnson's description of inertial-viscous effects.}.
These approximate expressions (denoted $JA$) of the two constitutive
functions, essentially are the result of:

1) an exact description of the high-frequency limits of the two functions,
density and compressibility, in terms of the concepts of ideal-fluid
tortuosity $\alpha _{\infty }$ and characteristic lengths $\Lambda $ \cite%
{joh} and $\Lambda ^{\prime }$ \cite{champ}, the viscous and thermal
relaxation processes being %\textquotedblleft frozen"
`frozen' in this limit\footnote{\label{frozen}In the frozen limit the
viscous and thermal relaxation processes have no time to develop; in the
relaxed limit they have enough time to fully develop.},

2) an exact description of the low-frequency limits, in terms of the
concepts of d.c. viscous and thermal permeabilities, $k_{0}$ \cite{darcy}
and $k_{0}^{\prime }$ \cite{lafargeth}, the viscous and thermal relaxation
processes being %\textquotedblleft relaxed"
`relaxed' in this limit, and finally,

3) an assumption that these frozen and relaxed limits 1) and 2) are
connected in the simplest reasonable manner, $i.e.$ by means of the simplest
`relaxation' functions of frequency $\omega $ having their singularities and
zeros lying on the imaginary half-axis in the complex $\omega $ plane (see
\cite{joh} Appendix A, and \cite{lafargeth} or \cite{lafarge} Appendix C).

That the singularities are on a half imaginary axis is a mathematical
expression of the fact that we actually restrict to the class of materials
allowing to apply the direct two-scale homogenization process. The pure
causality condition would only require the singularities to be in a half
plane. Our stronger assumption may be rephrased physically by saying that
the fluid velocity pattern at the pore scale is divergence-free for the
purpose of the determination of the density, and the pressure pattern is
uniform for the purpose of the determination of the compressibility; or in
essence, that the operators are local operators in space, at long
wavelengths, in the geometries we consider. The remaining assumption that
the functions are the simplest reasonable `relaxation' ones, consists in
making the additional but adjacent assumption that the considered geometries
manifest a relatively narrow -- not bimodal, for example -- distribution of
pore sizes. Then the distribution of poles on the imaginary axis is simple
and the whole pattern of response functions on the real axis is strongly
determined by the low- and high-frequency behaviors.

In this manner, a simple, resp. viscous and thermal, relaxation-transition
description of the density and compressibility functions is obtained, that
may be thought to be well-verified in a wide class of materials as long as
the wavelengths are large compared to the typical dimensions of the
coarse-graining averaging volumes. At this point, we may mention a similar
relaxation-transition approach, developed by Wilson \cite{wil1, wil2}. In
Wilson's approach, the emphasis is not made on the low-frequency and
high-frequency limits, but directly on the transition. The quality of the
description obtainable with Wilson's models is comparable to that of model $%
JA$, but the parameters become purely adjustable parameters not precisely
defined in terms of the microgeometry, and not clearly obtainable by
non-acoustical means.

Subsequently, the low-frequency relaxed limit 2) was made more precise by
Pride $et$ $al.$ \cite{pride} and Lafarge \cite{lafargeth} (the next viscous
and nontrivial thermal terms now being exactly described thanks to the
introduction of the additional notions of d.c. viscous and thermal
tortuosities $\alpha _{0}$ and $\alpha _{0}^{\prime }$ ), resulting in a
slightly improved (denoted $PL$) description of the relaxation transitions
of the two functions. Notice that some confusions were present in the
original works \cite{pride} and \cite{lafargeth} (not readily available) and
other subsequent ones (e.g. \cite{cort}), so that the presentation we give
later on, of the $PL$ description, may be worthwhile.

In Ref. \cite{fellah}, Fellah $et$ $al.$ concentrated on the time-domain
expression of the high-frequencies asymptotics implied by this model%
\footnote{\label{3f}A short, non exhaustive list of works concerned with the
time-domain description is: \cite{plyu}-\cite{b5}.}. Expanding, in the
high-frequency limit, the $PL$ density and compressibility in powers of the
inverse Stokes number $S_{T}^{-1}$ (defined as the ratio between boundary
layer thickness and characteristic pore size), and retaining the (exact)
zero and first order terms and the (model-dependent) second order terms,
they derived an asymptotic time-domain pressure wave equation. Using
fractional-derivative and Laplace-transform calculus, they were able to
solve this equation in elegant manner through the calculation of a
corresponding asymptotic Green's function of the unbounded medium. Recall
that a Green's function or impulse response, as a function of time, extends
and flattens when observed at fixed locations more and more remote from the
spatial point where it originates.

Now, the results of the calculations made in \cite{fellah} seem to indicate
that the terms of second order yield a significant effect on the Green's
function for resistive porous materials, and, in addition, that the effect
is mainly an effect on the amplitude without noticeable distortion of the
time wave pattern (see Figs. 2 and 3 of paper \cite{fellah}).

In the present paper, our purpose is threefold.

First, we wish taking advantage of recent clarifications regarding nonlocal
(spatial dispersion) effects\cite{laf1} to propose a lucid review of the
above local (frequency dispersion) theory, stressing the physical hypotheses
behind it. By recalling the construction principle of the models $JA$ and $%
PL $, we make clear that these models $a$ $priori$ lead to inaccurate
descriptions of the second order high-frequency terms $S_{T}^{-2}$. This is
highlighted on the simple example of cylindrical circular tubes. This is the
matter of section \ref{BE}.

Next, we give a very simple analytic derivation of the
attenuation-without-distortion finding of Ref. \cite{fellah} summarized
above. Indeed, generalizing to arbitrary geometry a work done by Polack $et$
$al.$ \cite{polack} for cylindrical circular tubes, we show the following
exact property (for 1D propagation along one axis $x$):
\begin{equation}
G_{o(2)}(x,t)=G_{o(1)}(x,t)\exp (-x/\mathcal{L}),  \label{O2}
\end{equation}%
where $G_{o(1)}(x,t)$ and $G_{o(2)}(x,t)$ are the two first asymptotic
Green's functions computed by retaining in the wavenumber the terms up to
the first and second order on the inverse Stokes number $S_{T}^{-1}$
respectively. This is done in section \ref{SW}, using a scaled-form of the
Green's function. The first is only determined by $\alpha _{\infty }$, $%
\Lambda $ and $\Lambda ^{\prime }$, and exactly predicted by the models $JA$
and $PL$. The second demonstrates the mentioned
attenuation-without-distortion effect through the exponential. The decay
length $\mathcal{L}$, however, depends in part of the next frozen
parameters, two viscous and thermal purely geometrical characteristic
surfaces $\Sigma $ and $\Sigma ^{\prime }$ involved in the above-mentioned
terms $S_{T}^{-2}$:
\begin{equation}
\mathcal{L}=\frac{2\Lambda ^{2}c_{f}}{\nu \sqrt{\alpha _{\infty }}}\left[
\frac{3\Lambda ^{2}}{\Sigma }-1+\frac{2(\gamma -1)\Lambda }{\Lambda ^{\prime
}\sqrt{\Pr }}+\frac{(\gamma -1)\Lambda ^{2}}{\Lambda ^{\prime 2}\Pr }\left(
\frac{3\Lambda ^{\prime 2}}{\Sigma ^{\prime }}-3-\gamma \right) \right]
^{-1},  \label{decay L}
\end{equation}%
where the fluid constants are, $\nu =\eta /\rho _{f}$ the fluid kinematic
viscosity, $c_{f}=\sqrt{K_{f}/\rho _{f}}$ the adiabatic speed of sound, $\Pr
$ the Prandtl number, and $\gamma =c_{p}/c_{v}$ the ratio of heat
coefficients. Now, because the models $JA$ and $PL$ (especially $JA$) give
inaccurate predictions for the $S_{T}^{-2}$ terms, this decay length $%
\mathcal{L}$ will not be accurately captured by the models. Indeed, on the
simple example of cylindrical circular tubes it can be checked that $%
\mathcal{L}$ is completely misrepresented by model $JA$, which gives a
negative estimate for it, and still largely underestimated by model $PL$,
which produces only about $1/3$ of its correct value, due to $50\%$
overestimation of surfaces $\Sigma $ and $\Sigma ^{\prime }$. Both models $%
JA $ and $PL$ do not describe the correct high-frequency
attenuation-without-distortion effect. Technical details are given in
Appendix to lighten the main text.

Finally, our last objective is to show that, in spite of its faulty
description of the decay length $\mathcal{L}$ (\ref{decay L}), the model $PL$%
, nevertheless furnishes a relatively precise description of the complete
exact Green's function, especially when compared to other formulas. The
asymptotic analytic Green's function $G_{o(2)}$ provides a reasonable
description of the complete Green's function with the same number of
parameters as $JA$, but it uses parameters $\Sigma $ and $\Sigma ^{\prime }$
which are unknown in general. The merits and drawbacks of the different
descriptions are illustrated on the example of circular pores -- $i$)
computing the exact Green's function through FFT and the known Zwikker and
Kosten full frequency formulas; $ii$) computing likewise the model Green's
functions $JA$ and $PL$ through FFT; $iii$) computing the exact asymptotic
Green's function $G_{o(2)}$ through (\ref{O2}-\ref{decay L}) with the known
exact values of the involved parameters (namely $\alpha _{\infty }=1$, $%
\Lambda =\Lambda ^{\prime }=R$, $\Sigma =\Sigma ^{^{\prime }}=R^{2}$); and $%
vi$) computing the `model asymptotic' Green's functions through (\ref{O2}-%
\ref{decay L}) with the model values of the involved parameters ($\alpha
_{\infty }=1$, $\Lambda =\Lambda ^{\prime }=R$, and $\Sigma =\Sigma ^{\prime
}=\frac{3}{2}R^{2}$ for $PL$ and $\Sigma =\Sigma ^{\prime }=\infty $ for $JA$%
). The results of FFT computation are given in section \ref{FFTC} for the
full-frequency models, while the results for asymptotic expansions are given
in Appendix.

\section{Basic equations\label{BE}}

We start by recalling the form of the macroscopic equivalent-fluid equations
in the frequency domain (see \cite{joh, lafarge, allard, bruneau}):%
\begin{equation}
\rho _{f}\alpha (\omega )i\omega v_{i}=\nabla _{i}p\text{ \ ; \ }%
K_{f}^{-1}\beta (\omega )i\omega p=\bm{\nabla }.v\text{\ ,}  \label{11}
\end{equation}%
where by definition, $\mathbf{v}$ and $p$ are the macroscopic velocity and
pressure obtained by coarse-graining (averaging) the microscopic fluid
velocity and pressure fields, $-i\omega $ is the time derivative, $\rho _{f}$
and $K_{f}$ are the saturating-fluid density and adiabatic bulk modulus, and
$\alpha (\omega )$ and $\beta (\omega )$ are the dynamic tortuosity and the
dynamic compressibility. Notice that for simplicity, isotropy or 1D
propagation along one principal axis is assumed, so that $\alpha (\omega )$
is a scalar.

Eqs. (\ref{11}) make apparent the frequencies but not the wavenumbers
associated to the time- and space-variable fields. This is consistent with
the hypothesis that the effects of spatial dispersion are negligible. A
necessary -- but not sufficient -- condition for this, is that the
wavelengths are large. The complementary condition which will ensure that
spatial locality is verified, is that no very different pore sizes are
present, so that in particular, the presence of structures in the form of
Helmholtz resonators is excluded. Indeed, with resonators, the fluid
exchanged to and fro, is always associated by mass conservation with a
corresponding spatial inhomogeneity in the wavefield. The resonance cannot
occur without concomitant spatial inhomogeneity in the macroscopic fields.
As such, it is an effect of spatial dispersion, see \cite{landauel} p. 360,
no matter how large the wavelengths actually are. Now, if we restrict
sufficiently the possible geometries, resonances cannot occur and the
long-wavelength propagation becomes practically devoid of spatial dispersion
$i.e.$ is given by the traditional homogenization.

The application of the asymptotic two-scale homogenization then yields two
different microscopic boundary value formal problems to be solved for
determining the two constitutive functions $\alpha(\omega)$ and $%
\beta(\omega)$ \cite{lafarge}.

The first action-response problem specifies the (velocity) response of a
viscous incompressible fluid subject to an applied time harmonic, spatially
uniform bulk force source term:
\begin{eqnarray}
\frac{-i\omega \rho _{f}}{\eta }\mathbf{w} &=&-\bm{\nabla }\Pi +\boldsymbol{%
\nabla }^{2}\mathbf{w}+\mathbf{e},  \label{d1} \\
\bm{\nabla }\cdot \mathbf{w} &=&0,  \label{d2}
\end{eqnarray}%
in the pore space (with $\Pi $ a periodic or stationary random field, in
periodic or stationary random geometries), and satisfying on the pore
surface (no slip condition),
\begin{equation}
\mathbf{w}=\mathbf{0}.  \label{d3}
\end{equation}%
In this problem, $\mathbf{e}$ is a dimensionless unit vector in the
direction of the applied bulk force, and $\mathbf{w}$ is the scaled response
velocity field (dimension of $\mathit{length}^{2}$). It determines the
dynamic permeability $k(\omega )$ and dynamic tortuosity $\alpha (\omega )$
introduced in the landmark paper by Johnson $et$ $al.$ \cite{joh}, by the
relations:
\begin{equation}
k(\omega )=\frac{\eta \phi }{-i\omega \rho _{f}\alpha (\omega )}=\langle
\mathbf{w}\rangle \cdot \mathbf{e},  \label{d4}
\end{equation}%
where $\langle \rangle $ is the coarse-graining averaging operation in the
pore space, and $\phi $ is the porosity (specific connected pore volume).

The second action-response problem specifies the (excess temperature)
response of a thermal fluid subject to an applied time harmonic, spatially
uniform pressure source term:
\begin{equation}
\frac{-i\omega \Pr \rho _{f}}{\eta }\theta =\bm{\nabla }^{2}\theta +1,
\label{d1'}
\end{equation}%
in the pore space, and satisfying on the pore surface (no temperature-jump
condition\footnote{%
The solid specific volume ($1-\phi$) is assumed sufficiently important to
ensure that the specific fluid-solid mass ratio is small; then the solid is
inert thermally and remains at ambient temperature.}),
\begin{equation}
\theta =0.  \label{d3'}
\end{equation}%
Here, $1$ is a dimensionless unit constant representing the applied
pressure, and $\theta $ is the scaled response excess temperature field
(dimension of $\mathit{length}^{2}$). It determines the functions $k^{\prime
}(\omega )$ and $\alpha ^{\prime }(\omega )$, thermal counterparts of
functions $k(\omega )$ and $\alpha (\omega )$ introduced by Lafarge \cite%
{lafargeth,lafarge}, and then, the effective compressibility $\beta (\omega
) $, by the following relationships:
\begin{equation}
k^{\prime }(\omega )=\frac{\eta \phi }{-i\omega \Pr \rho _{f}\alpha ^{\prime
}(\omega )}=\langle \theta \rangle \text{ \ ; \ }\beta (\omega )=\gamma -%
\frac{\gamma -1}{\alpha ^{\prime }(\omega )}  \label{d4'}
\end{equation}%
where $\gamma $ is the ratio of the specific heats.

The above action-response problems and response-factor identifications (\ref%
{d4}) and (\ref{d4'}) are written in blind manner by applying the
traditional two-scale asymptotic homogenization. As we have insisted, it
makes the important assumption that it is possible to neglect spatial
dispersion. This is manifested in the first problem by the force source term
$\mathbf{e}$ which is set to a spatial constant (simultaneously, the
velocity field $\mathbf{v}$ is represented by a divergence-free field), and
in the second problem by the pressure source term $1$ which is also set to a
spatial constant (gradient-free)\footnote{\label{5f}In fact, what is really
meant here is that in this second problem, locally, the pressure field may
be viewed as having a uniform gradient; but the pressure linear variation
around the mean may be omitted in a coarse graining volume, as it has no
effect on the mean temperature. As soon as spatial dispersion is introduced,
it no longer makes sense to represent the source terms by spatial constants.
The spatial inhomogeneity of the source terms must be considered.}. Indeed,
we could have written these problems and identifications directly without
using the homogenization process, on the sole basis of assuming that spatial
dispersion effects are absent. Now, as a result of the divergence-free and
gradient-free nature of velocity and pressure in the given action-response
problems, it can be shown that the functions $k(\omega)$ and $k^{\prime
}(\omega)$ have purely imaginary singularities \cite{joh}, \cite{lafarge},
and thus are smooth functions on the real axis -- this is the point 3)
mentioned in Introduction. This crucial point in the construction of the
models will be considered at more length in section $2.2$.

\subsection{Frozen and relaxed limits}

In absence of a complete information on the microgeometry, the two
above-mentioned microscopic formal boundary-value problems cannot entirely
be worked out and their exact detailed solutions (from which $\alpha (\omega
)$ and $\beta (\omega )$ can in principle be extracted by coarse-graining,
see Eqs.(\ref{d4}) and (\ref{d4'})) are missing. Nevertheless, in the limit
of high-frequencies and low-frequencies, some general characteristics of the
solutions and corresponding functions $\alpha (\omega )$ and $\beta (\omega
) $ may be obtained. That may be sketched as follows.

\subsubsection{High frequencies (frozen limit)}

In the limit of high frequencies, the viscous and thermal terms $\bm{\nabla }%
^{2}\mathbf{w}$ and $\bm{\nabla }^{2}\theta $ become negligibly small
compared to the other terms. The fluid motions, except for vanishingly small
viscous and thermal boundary layers at the pore walls, become close to those
of an inviscid nonconducting fluid ($\eta =0$, $\kappa =0$, with $\kappa $
the thermal conductivity $c_p \eta/\Pr)$. We refer to this limit simply as
the `frozen limit' (see footnote \ref{frozen}). Accordingly, the quantities $%
\frac{-i\omega \rho _{f}}{\eta }\mathbf{w}$ and $\frac{-i\omega \Pr \rho _{f}%
}{\eta }\theta $ everywhere tend (except at the pore walls) to the
%\textquotedblleft frozen"
`frozen' fields $\mathbf{E}$ and $I$ verifying the following equations in
the pore space, (with $\varphi $ a periodic or stationary random field, in
periodic or stationary random geometries),
\begin{eqnarray}
\mathbf{E} &=&-\bm{\nabla }\varphi +\mathbf{e}\text{ \ ; \ }I=1,  \label{dp1}
\\
\bm{\nabla }\cdot \mathbf{E} &=&0,  \label{dp2}
\end{eqnarray}%
and verifying at the pore walls ($\mathbf{n}$ is the normal on the latter),
\begin{equation}
\mathbf{E}\cdot \mathbf{n}=\mathbf{0}.  \label{dp3}
\end{equation}

Now assume, following Johnson $et$ $al.$ \cite{joh} and Allard \cite{allard}%
, that the pore-surface interface appears locally plane in this asymptotic
high-frequency frozen limit. This is in principle an assumption that the
viscous and thermal boundary layer thicknesses $\delta =\sqrt{\frac{2\eta }{%
\rho _{f}\omega }}$ and $\delta ^{\prime }=\sqrt{\frac{2\eta }{\rho
_{f}\omega \Pr }}$ eventually become small compared to a characteristic
radius of curvature of the pore surface\footnote{%
In practice this is not strictly necessary: because of the smooth nature of
response functions, the limiting behaviors (\ref{hf}) become meaningful much
more rapidly.}. Then, the functions $\alpha (\omega )$ and $\alpha ^{\prime
}(\omega )$ expand in integral power series of these thicknesses, which
allows us writing $a$ $priori$\footnote{\label{fract}Notice that the case of
fractal geometry which modifies the exponent $1/2$ in the first correction
terms -- see \cite{joh} -- is excluded by the assumption that the pore walls
appear locally flat at the scale of the boundary layer thickness; the
presence of sharp edges which modifies the exponent $1$ in the second
correction terms -- see \cite{cortis} -- is also excluded by this assumption.%
},
\begin{eqnarray}
\alpha (\omega ) &=&\alpha _{\infty }+\frac{2\alpha _{\infty }}{\Lambda }%
\left( \frac{\eta }{-i\omega \rho _{f}}\right) ^{1/2}+\frac{3\alpha _{\infty
}}{\Sigma }\left( \frac{\eta }{-i\omega \rho _{f}}\right) +O\left( \frac{1}{%
-i\omega }\right) ^{3/2}\text{ \ ; \ }  \notag \\
\alpha ^{\prime }(\omega ) &=&\alpha _{\infty }^{\prime }+\frac{2\alpha
_{\infty }^{\prime }}{\Lambda ^{\prime }}\left( \frac{\eta }{-i\omega \rho
_{f}\Pr }\right) ^{1/2}+\frac{3\alpha _{\infty }^{\prime }}{\Sigma ^{\prime }%
}\left( \frac{\eta }{-i\omega \rho _{f}\Pr }\right) +O\left( \frac{1}{%
-i\omega }\right) ^{3/2}.  \label{hf}
\end{eqnarray}%
The geometrical parameters $\alpha _{\infty }$ and $\alpha _{\infty
}^{\prime }$ (dimensionless) and $\Lambda $ and $\Lambda ^{\prime }$
(dimension of $\mathit{length}$) must be some pore averages constructed with
the frozen fields $\mathbf{E}$ and $I$. The next order geometrical
parameters $\Sigma $ and $\Sigma ^{\prime }$ (dimension of $\mathit{surface}$%
) are dependent on other fields involved in the asymptotic limit and have
not been worked out so far. Detailed calculations made by Johnson and Allard
\cite{joh, champ}, to which we refer the reader, show that the parameters $%
\alpha _{\infty }$, $\alpha _{\infty }^{\prime }$, $\Lambda $ and $\Lambda
^{\prime }$ may be written as follows (see also, in the most detailed manner
for $\Lambda $, \cite{cortis}):
\begin{equation}
\frac{1}{\alpha _{\infty }}=\left\langle \mathbf{E}\right\rangle \cdot
\mathbf{e}=\frac{\left\langle \mathbf{E}\right\rangle ^{2}}{\left\langle
\mathbf{E}^{2}\right\rangle }\text{ \ ; \ }\frac{1}{\alpha _{\infty
}^{\prime }}=\left\langle \mathbf{I}\right\rangle =1,
\end{equation}%
and,
\begin{equation}
\frac{2}{\Lambda }=\frac{\int_{S_{p}}\mathbf{E}^{2}dS}{\int_{V_{f}}\mathbf{E}%
^{2}dV}\text{ \ ; \ }\frac{2}{\Lambda ^{\prime }}=\frac{\int_{S_{p}}I^{2}dS}{%
\int_{V_{f}}I^{2}dV}=\frac{\int_{S_{p}}dS}{\int_{V_{f}}dV}\text{ \ ,}
\label{lambd}
\end{equation}%
where $S_{p}$ denotes the pore walls and $V_{f}$ denotes the connected pore
volume.

Parameter $\alpha_{\infty}/\phi$ is Brown's electric formation factor \cite%
{brown} ($\mathbf{E}$ represents either an inviscid-fluid scaled velocity or
acceleration field for the problem of incompressible inviscid fluid flow,
accelerating under the action of external bulk force or applied pressure
drop, or else, an electric scaled field for the problem of electrical
conduction in the bulk fluid \cite{avell}, \cite{lafcom}). Parameter $%
\Lambda $ is an effective pore radius for dynamically connected pore sizes
which was introduced by Johnson $et$ $al.$ \cite{johks} for the problem of
electrical conduction in the bulk fluid, perturbed by a thin, different
conducting layer at the pore walls. Parameter $\Lambda ^{\prime }$ is a
length characterizing a simpler effective pore radius -- twice the
fluid-volume to fluid-surface ratio -- sometimes referred to as the
Kozeny-Carman radius; Allard \cite{champ} identified it as the thermal
counterpart of parameter $\Lambda $. An incomplete reasoning to obtain the
parameter $\Lambda $, leading to an incomplete expression $%
2/\Lambda=\int_{S_{p}}\mathbf{E}\cdot\mathbf{e}\, dS/\int_{V_{f}}\mathbf{E}%
^{2}dV$, is often made, e.g. \cite{zhou} \cite{auri2} \cite{pridecoupling};
the reasoning inaccuracy\footnote{%
The ignorance, in the bulk, $i.e.$ outside the viscous boundary layer or the
perturbed conducting layer, of a perturbation contribution due to a
perturbed ideal-fluid or electrical bulk flow field orthogonal to the
leading bulk flow field $\mathbf{E}$, and having, contrary to the latter,
nonzero normal components at the pore walls.} was clarified and corrected in
\cite{cortis} along a line tentatively sketched in \cite{avell} (Appendix D).

\subsubsection{Low frequencies (relaxed limit)}

In the opposite relaxed limit of low frequencies, the viscous and thermal
terms $\boldsymbol{\nabla }^{2}\mathbf{w}$ and $\boldsymbol{\nabla }%
^{2}\theta $ eventually become much greater than the inertial terms $\frac{%
-i\omega \rho _{f}}{\eta }\mathbf{w}$ and $\frac{-i\omega \Pr \rho _{f}}{%
\eta }\theta $ and the boundary layers extend to the whole fluid.
Accordingly, the fields $\mathbf{w}$ and $\theta $ everywhere tend to the
d.c. -- or %\textquotedblleft relaxed"
`relaxed' -- velocity and excess temperature fields $\mathbf{w}_{0}$ and $%
\theta _{0}$ verifying, in the pore space (with $\Pi _{0}$ a periodic or
stationary random field, in periodic or stationary random geometries),
\begin{eqnarray}
\boldsymbol{\nabla }^{2}\mathbf{w}_{0} &=&\boldsymbol{\nabla }\Pi _{0}-%
\mathbf{e}\text{ \ ; \ }\boldsymbol{\nabla }^{2}\theta _{0}=-1,  \label{dv1}
\\
\boldsymbol{\nabla }\cdot \mathbf{w}_{0} &=&0,  \label{dv2}
\end{eqnarray}%
and verifying, at the pore walls,
\begin{equation*}
\mathbf{w}_{0}=\mathbf{0}\text{ \ ; \ }\theta _{0}=0.
\end{equation*}%
In this limit the functions $\alpha (\omega )$ and $\alpha ^{\prime }(\omega
)$ expand in Laurent's series,
\begin{equation}
\alpha (\omega )=\frac{\eta \phi }{-i\omega \rho _{f}k_{0}}+\alpha
_{0}+O(-i\omega )\text{ \ ; \ }\alpha ^{\prime }(\omega )=\frac{\eta \phi }{%
-i\omega \rho _{f}\Pr k_{0}^{\prime }}+\alpha _{0}^{\prime }+O(-i\omega ).
\label{bf}
\end{equation}%
The two first intervening geometrical parameters, $k_{0}$ (Darcy's viscous
permeability) and $k_{0}^{\prime }$ (its thermal counterpart \cite%
{lafargeth, lafarge}) on one hand, both having dimension of $\mathit{length}%
^{2}$, and $\alpha _{0}$ (viscous tortuosity) and $\alpha _{0}^{\prime }$
(its thermal counterpart) on the other hand, both being dimensionless, are
pore averages constructed with the relaxed fields $\mathbf{w}_{0}$ and $%
\theta _{0}$. Indeed, simple calculations show that they may be written \cite%
{norr,lafargeth},
\begin{eqnarray}
k_{0} &=&\phi \left\langle \mathbf{w}_{0}\right\rangle \cdot \mathbf{e}\text{
\ ; \ }k_{0}^{\prime }=\phi \left\langle \theta _{0}\right\rangle ,
\label{lfp1} \\
\alpha _{0} &=&\frac{\left\langle \mathbf{w}_{0}^{2}\right\rangle }{%
\left\langle \mathbf{w}_{0}\right\rangle ^{2}}\text{ \ ; \ }\alpha
_{0}^{\prime }=\frac{\left\langle \theta _{0}^{2}\right\rangle }{%
\left\langle \theta _{0}\right\rangle ^{2}}.  \label{lfp2}
\end{eqnarray}%
Mention that, among other things it is possible to show that \cite{torq,
brown, norr, lafargeth}, whatever the geometry,
\begin{equation}
\alpha _{0}>\alpha _{\infty }\geq 1,
\end{equation}%
\begin{equation}
k_{0}\leq k_{0}^{\prime }\text{ \ ; \ }\alpha _{0}\geq \alpha _{0}^{\prime
}>1\text{ \ ; \ }\Lambda \leq \Lambda ^{\prime },
\end{equation}%
the equalities being satisfied only for the case of aligned cylindrical
pores.

\subsection{Full-frequency models\label{ffm}}

As explained before, it is only for sufficiently simple geometries that the
assumed long-wavelength nature of the considered fields, automatically imply
that the microscopic flow-field can be considered divergence-free for the
purpose of determining the density (Eq. (\ref{d2})), and likewise, the
excess pressure field can be considered gradient-free for the purpose of
determining the compressibility.

Now, Johnson $et$ $al.$ have made the important observation that, because
the velocity field is locally divergence-free, the singularities of
functions $\alpha(\omega )$ and $k(\omega )$ -- poles, zeros, and branch
points -- necessarily are purely imaginary (see \cite{joh} Appendix A). This
characteristics of the response functions is explicitly apparent in
Avellaneda and Torquato's solution of principle of the problem (\ref{d1}-\ref%
{d3}), written in terms of the Stokes operator's eigenmodes and relaxation
times \cite{avell}.

For the functions $\alpha^{\prime }(\omega)$ and $k^{\prime }(\omega)$, it
was similarly shown \cite{lafargeth} that, because the excess pressure field
can be considered gradient-free\footnote{\label{8f}See footnote \ref{5f}},
the singularities of functions $\alpha^{\prime }(\omega )$ and $k^{\prime
}(\omega )$ also are purely imaginary (see \cite{lafarge} Appendix C).
Again, this characteristics of the response functions is explicitly apparent
in a solution of principle of the problem (\ref{d1'}-\ref{d3'}), written in
terms of the Laplace operator's eigenmodes and relaxation times \cite%
{lafargeth}, in exactly the same manner as \cite{avell}.

The wanted functions must therefore have simple smooth behaviors on the real
axis. Indeed, the DRT (distribution of relaxation times) formalism of
Avellaneda and Torquato may be used to explicitly show that in the Laplace
domain ($s=-i\omega >0$) the hydrodynamic drag function $\lambda(s)=1-1/%
\alpha(s)$ is always a strictly decreasing positive function on the real
axis $s>0$, and the same holds true in the same manner for the corresponding
thermal function $\lambda^{\prime }(s)=1-1/\alpha^{\prime }(s)$ -- see \cite%
{lafargeth}. Alternatively, more recently, the strictly decreasing nature of
functions $\Re \alpha(\omega)$ and $\omega \Im \alpha(\omega)$ versus real
frequency has been shown in elegant manner by a variational formulation
making use of the divergence-free nature (\ref{d2}) of the microscopic flow
fields \cite{bouting}.

Finally, as observed by Johnson $et$ $al.$ \cite{joh}, due to the special
location of their singularities and the adjacent hypothesis of the absence
of very different pore sizes, the functions $k(\omega)$ and $\alpha(\omega)$
as well as the functions $k^{\prime }(\omega)$ and $\alpha^{\prime }(\omega)$%
, may be seeked as the simplest ones satisfying both the frozen and relaxed
limits.

To proceed, let us introduce a Stokes number constructed using Johnson's
concept of dynamically connected pore size $\Lambda $, $viz$:
\begin{equation}
S_{T}=\Lambda \sqrt{\frac{-i\omega \rho _{f}}{\eta }}.  \label{12}
\end{equation}%
Johnson has proposed the following expression of dynamic tortuosity $\alpha
(\omega )$,
\begin{equation}
\alpha (\omega )=\alpha _{\infty }\left( 1+\frac{8}{MS_{T}^{2}}\sqrt{1+\frac{%
M^{2}}{16}S_{T}^{2}}\right) ,  \label{j14}
\end{equation}%
where
\begin{equation}
M=\frac{8\alpha _{\infty }k_{0}}{\phi \Lambda ^{2}},  \label{j15a}
\end{equation}%
is a dimensionless shape factor associated to the geometry. This expression
is the simplest analytical ansatz that yields the exact first two terms at
high frequencies (Eq. (\ref{hf})) and the exact first leading term at low
frequencies (Eq. (\ref{bf})), and automatically satisfies the condition on
singularities. As such, and for the general reasons discussed before, it is
expected to provide a reasonable description of the exact function $%
\alpha(\omega)$. Factors of 8 are introduced for convenience in Eqs. (\ref%
{j14}-\ref{j15a}), so that $M=1$ for cylindrical circular pores.

Similarly, to describe the function $\alpha ^{\prime }(\omega )$ let us
introduce a second Stokes number corresponding to thermal effects:%
\begin{equation}
S_{T}^{\prime }=\Lambda ^{\prime }\sqrt{\frac{-i\omega \rho _{f}\Pr }{\eta }}%
.  \label{13}
\end{equation}%
The ratio $S_{T}^{\prime }/S_{T}=\sqrt{\Pr }$ $\Lambda ^{\prime }/\Lambda $
\ is of the order of $\Pr^{1/2}$, not far from unity. Proceeding as did
Johnson, Lafarge \cite{lafargeth} proposed to write,
\begin{equation}
\alpha ^{\prime }(\omega )=1+\frac{8}{M^{\prime }S_{T}^{\prime 2}}\sqrt{1+%
\frac{M^{\prime 2}}{16}S_{T}^{\prime 2}},  \label{l14p}
\end{equation}%
where
\begin{equation}
M^{\prime }=\frac{8k_{0}^{\prime }}{\phi \Lambda ^{\prime 2}},  \label{l15a}
\end{equation}%
is the thermal counterpart of shape factor $M$ ($M^{\prime }=1$ for
cylindrical circular pores), giving in turn a definite model for the dynamic
compressibility%
\begin{equation}
\beta (\omega )=\gamma -(\gamma -1)/\alpha ^{\prime }(\omega ).  \label{15}
\end{equation}%
As this model is the elaboration of Allard's original attempt to transpose
Johnson's modelling to thermal effects \cite{champ}, we refer to the
combined modeling of functions $\alpha $ and $\beta $ Eqs. (\ref{12}--\ref%
{15}) as to Johnson-Allard's ($JA$).

Subsequently, Pride $et$ $al.$ \cite{pride}, while studying oscillating
viscous flow in convergent-divergent channels, found that the simple formula
(\ref{j14}) may significantly underestimate the imaginary part of dynamic
permeability $k(\omega)$ at low frequencies. To remedy this, they proposed
different modified formulas. In essence these are formulas capable to
account for the exact value of parameter $\alpha _{0}$, which, in such
channels, may be significantly increased as compared to Johnson's.
Notwithstanding, this parameter $\alpha _{0}$ is not singled out in \cite%
{pride}. Its identification by Eq. (\ref{lfp2}) (see \cite{norr} and \cite%
{lafargeth}) shows that it is constructed like the tortuosity $\alpha_\infty$%
, this time for the `Poiseuille'-like velocity pattern. Therefore it is a
measure of %\textquotedblleft disorder"
`disorder' of the %\textquotedblleft Poiseuille"
`Poiseuille' flow, and as such it is increased not only by the
convergent-divergent mechanism considered by Pride $et$ $al.$, but also, $%
e.g.$, by irregularities in the distribution of solid inclusions leading to
the existence of privileged flow paths. Whatever the cause of the enhanced
%\textquotedblleftPoiseuille disorder"
`Poiseuille disorder', a significant increase of factor $\alpha_0$ as
compared to Johnson's value $\alpha_0=\alpha_{\infty}(1+\frac{M}{4})$, will
make it more necessary to modify Johnson's formula. Similar formal
considerations hold true also, $mutatis$ $mutandis$, for thermal effects and
the function $\alpha ^{\prime }$. Here also, a significant increase of $%
\alpha^{\prime }_{0}$ may result when replacing a regular distribution of
solid inclusions by an irregular one (which leaves unchanged the thermal
characteristic length $\Lambda^{\prime }$).

Now, among the different modifications proposed in \cite{pride} the first
was the simplest one, capable to yield the exact first two terms at high and
low frequencies, and simultaneously, to automatically fulfil the condition
on singularities whatever the values of parameters $\phi $, $k_{0} $, $%
\alpha _{0}$, $\Lambda $, and $\alpha _{\infty }$\footnote{%
Incidentally, this nice feature of the formula was missed in \cite{pride}
and \cite{lafargeth}: in \cite{pride} there are mistaken considerations
concerning the formula, corrected in \cite{lafargeth} but still with
mistaken considerations on the singularities, repeated in \cite{cort}.}.
Finally, this formula was expressed by Lafarge in terms of the parameter $%
\alpha _{0}$ and the same description was then immediately transferable to
thermal effects. The corresponding Pride-Lafarge's ($PL$) model formulas are:%
\begin{eqnarray}
\alpha (\omega ) &=&\alpha _{\infty }\left( 1+\frac{8}{MS_{T}^{2}}\left(
1-q+q\sqrt{1+\frac{M^{2}}{16q^{2}}S_{T}^{2}}\right) \right) ,  \label{14} \\
\alpha ^{\prime }(\omega ) &=&1+\frac{8}{M^{\prime }S_{T}^{\prime 2}}\left(
1-q^{\prime }+q^{\prime }\sqrt{1+\frac{M^{\prime 2}}{16q^{\prime }{}^{2}}%
S_{T}^{\prime 2}}\right) ,  \label{14p} \\
\beta (\omega ) &=&\gamma -(\gamma -1)/\alpha ^{\prime }(\omega ),
\label{be}
\end{eqnarray}%
where $M$ and $M^{\prime }$ are as before, and $q$ and $q^{\prime }$ are the
new shape factors given by,
\begin{equation}
q=\frac{1}{\alpha _{0}-\alpha _{\infty }}\frac{2k_{0}\alpha _{\infty }^{2}}{%
\phi \Lambda ^{2}}\text{ \ ; \ }q^{\prime }=\frac{1}{\alpha _{0}^{\prime }-1}%
\frac{2k_{0}^{\prime }}{\phi \Lambda ^{\prime 2}}.  \label{15b}
\end{equation}

These expressions reduce to $JA $'s by setting $q=q^{\prime}=1$ and are
simple transformations of the latter: they apply the simple group
transformation $F(S_T) \to 1-q+q F(S_{T}/q)$ to the basic Johnson's square
root function $F_{J}(S_{T})=\sqrt{1+\frac{M^2}{16}S_{T}^{2}}$. Two
successive applications of the transformation (with parameters $q$ and $p$)
yield another same transformation (with parameter $qp$) which preserves both
low- and high-frequency limits $F(S_T) \to 1 + O(S_T^2)$ and $F(S_T) \to
\frac{M}{4}S_T$. In this way there is some unicity in the $PL$ modification,
which is not extendible in obvious very simple manner.

It must be realized that, as a side result of the strong constraints imposed
by the special location of singularities and as long as the geometry is
relatively simple, the $PL$ expressions constructed with one more exact term
than $JA$'s at low frequencies, will also describe in a slightly more
accurate manner all of the viscous and thermal relaxation\footnote{%
The incorrect statement that the $PL$ description essentially improves the
low frequencies if often made in literature. In reality because of the pole
in $\alpha(\omega)$ and the way $\beta(\omega)$ is related to $%
\alpha^{\prime }(\omega)$ -- see (\ref{be}), the small departures between $%
PL $ and $JA$ are mainly perceptible in the region of intermediate
frequencies.}. Nevertheless, it should not be hoped that it is possible to
gain, by means of this description, very meaningful information on the
frozen parameters $\Sigma$ and $\Sigma^{\prime }$.

If the model expressions (\ref{14}-\ref{15b}) were exact, the comparison of
their high-frequency expansions with the exact ones (\ref{hf}) would imply

\begin{equation}  \label{qPridem}
q=1-\frac{3\alpha_{\infty}k_0}{\phi \Sigma}\text{ \ ; \ }q^{\prime }=1-\frac{%
3k^{\prime }_0}{\phi \Sigma^{\prime }},
\end{equation}
hence giving a fixed relation between the set of relaxed and frozen
parameters. But the expressions (\ref{14}-\ref{14p}) are not exact, and the
numbers $q$ and $q^{\prime }$ in (\ref{15b}-\ref{qPridem}), obtained by low-
or high-frequency matching of these non-exact expressions, will not be the
same in general. Thus we should not hope that the high-frequency expansions
obtained with $PL$ model:
\begin{eqnarray}
\alpha (\omega ) &=&\alpha _{\infty }\left( 1+\frac{2}{S_{T}}+\frac{8(1-q)}{%
MS_{T}^{2}}+O\left( \frac{1}{S_{T}^{3}}\right) \right) ,  \label{16} \\
\beta (\omega ) &=&1+(\gamma -1)\left( \frac{2}{S_{T}^{\prime }}+\frac{%
8(1-q^{\prime })}{M^{\prime }S_{T}^{\prime }{}^{2}}-\frac{4}{S_{T}^{\prime 2}%
}+O\left( \frac{1}{S_{T}^{\prime 3}}\right) \right) ,  \label{18}
\end{eqnarray}%
yield anything precise for the $O(2)$ terms. The example of cylindrical
circular tubes may serve to illustrate this in quantitative manner.

\subsection{\protect\bigskip The case of cylindrical pores\label{ccp}}

For a material with cylindrical circular pores of identical radius $R$ (say
for simplicity, all parallel and aligned along the direction of
propagation), the two boundary value problems Eqs. (\ref{d1}--\ref{d3}) and (%
\ref{d1'}--\ref{d3'}) determining $\alpha (\omega )$ and $\beta (\omega )$,
are easily entirely stated and solved. In effect, these are nothing but the
problems considered by Zwikker and Kosten \cite{zk} in simplifying (on
account of the wide separation between wavelength and tube radius) the
governing equations of the full Kirchhoff's theory of sound propagation in a
cylindrical circular tube \cite{kirch}. We may say, in this respect, that
the conventional equivalent-fluid theory neglecting spatial dispersion and
expressed by Eqs. (\ref{11}--\ref{d4'}), is the direct generalization to the
case of arbitrary geometry, of Zwikker and Kosten's classic theory. Now,
Zwikker and Kosten's result is that $\alpha (\omega )$, $\alpha^{\prime
}(\omega)$ and $\beta (\omega )$ express via Bessel functions as follows:
\begin{equation}
\frac{1}{\alpha (\omega )}=1-\chi (\omega )\text{ \ ; \ }\frac{1}{%
\alpha^{\prime }(\omega )}=1-\chi (\omega\Pr )\text{ \ ; \ }\beta (\omega
)=1+(\gamma -1)\chi (\omega \Pr ),  \label{ZW1}
\end{equation}%
where $\chi (\omega )$ is the following relaxation function,
\begin{equation}
\chi (\omega )=\frac{2J_{1}\left( (\frac{i\omega \rho _{f}}{\eta }%
R^{2})^{1/2}\right) }{(\frac{i\omega \rho _{f}}{\eta }R^{2})^{1/2}J_{0}%
\left( (\frac{i\omega \rho _{f}}{\eta }R^{2})^{1/2}\right) },  \label{ZW2}
\end{equation}%
(smoothly varying, in a sort `Davidson-Cole' pattern, from relaxed value 1
at low frequencies to frozen value 0 at high frequencies). Using the known
small-arguments and large-arguments series and asymptotic expansions of
Bessel functions (or Kelvin functions), it is simple to derive the following
low-frequency and high-frequency exact behaviors:

Low frequencies:
\begin{equation}
\alpha (\omega )=\frac{8\eta}{-i\omega \rho _{f}R^2}+\frac{4}{3}+O(-i\omega )%
\text{ \ ; \ }\alpha ^{\prime }(\omega )=\frac{8\eta}{-i\omega \rho _{f}\Pr
R^2}+\frac{4}{3}+O(-i\omega )  \label{LFc}
\end{equation}

High frequencies:
\begin{eqnarray}
\alpha (\omega ) &=&1+\frac{2}{R }\left( \frac{\eta }{-i\omega \rho _{f}}%
\right) ^{1/2}+\frac{3}{R^2 }\left( \frac{\eta }{-i\omega \rho _{f}}%
\right)+O\left( \frac{1}{-i\omega }\right)^{3/2} \text{ \ ; \ }  \notag \\
\alpha ^{\prime }(\omega ) &=&1+\frac{2}{R}\left( \frac{\eta }{-i\omega \rho
_{f}\Pr }\right) ^{1/2}+\frac{3}{R^2}\left( \frac{\eta }{-i\omega \rho
_{f}\Pr }\right)+O\left( \frac{1}{-i\omega }\right)^{3/2} .  \label{HFc}
\end{eqnarray}%
Then comparing these results with the general low- and high-frequency
expansions Eqs. (\ref{bf}) and (\ref{hf}), the following parameters values
are easily obtained:
\begin{equation}
k_{0}=k_{0}^{\prime }=\phi R^{2}/8\text{ \ ; }\alpha _{0}=\alpha
_{0}^{\prime }=4/3\text{ \ ; } \Sigma =\Sigma ^{\prime }=R^2 \text{ \ ; \ }
\Lambda =\Lambda ^{\prime }=R \text{ \ ; \ } \alpha _{\infty }=1.  \label{19}
\end{equation}

From Eqs. (\ref{15}) and (\ref{15b}), the circular-tube shape factors $M$, $%
q $ and $M^{\prime }$, $q^{\prime }$ are identified as:%
\begin{equation}
M=M^{\prime }=1\text{ \ ; \ }q=q^{\prime }=3/4,  \label{19b}
\end{equation}%
whereas for the modified $PL$ model (\ref{qPridem}) the latter are:
\begin{equation}
q=q^{\prime }=5/8.  \label{19bm}
\end{equation}

Let us now examine how far the models are consistent with the limits (\ref%
{LFc}-\ref{HFc}) and collection of parameters (\ref{19}).

When Johnson's values $q=q^{\prime }=1$ are used in the model expressions (%
\ref{14}-\ref{14p}), no $S_{T}^{2}=O(-1/i\omega)$ terms appear in the
high-frequency limit (\ref{hf}): the characteristic surfaces $\Sigma$ and $%
\Sigma^{\prime }$ predicted by model $JA$ are given infinite values whatever
the geometry (this can be seen also on Eqs. (\ref{qPridem})).
Simultaneously, in the low-frequency limit, the relaxed parameters $\alpha_0$%
, $\alpha^{\prime }_0$ in (\ref{bf}) are given as $\alpha_0=\alpha_\infty
\left(1+\frac{M}{4}\right)$, $\alpha^{\prime }_0= \left(1+\frac{M^{\prime }}{%
4}\right)$. Consider specifically the increments $\alpha_0-\alpha_\infty$
and $\alpha^{\prime }_0-1$, $i.e.$ the differences $[\alpha]_{relaxed}-[%
\alpha]_{frozen}$ produced by the viscous and thermal relaxation processes.
For the present case of cylindrical circular tubes, it follows that the
relative error $\left[()_{model}-()_{exact}\right]/ ()_{exact}$ made by the
model $JA$ on these increments, is a $(1/4-1/3)/(1/3)$ = -25\% error.

When the $PL$ $q$ and $q^{\prime }$ values (\ref{19b}) ($i.e.$ (\ref{15b}))
are used, the latter increments are exactly described, but a 50\%
overestimation still exists for the characteristic surfaces: putting the
values (\ref{19b}) in (\ref{qPridem}) yields $\Sigma=\Sigma^{\prime 2}/2$.

When the modified $PL$ $q$ and $q^{\prime }$ values (\ref{19bm}) ($i.e.$ (%
\ref{qPridem})) are used, the characteristic surfaces are exactly described
but there remain now a $(8/(4\text{x} 5) -1/3)/(1/3)$ = +20\% error on the
viscous and thermal increments.

Finally, when the values of $q$ and $q^{\prime }$ are taken as the
arithmetic mean of the $PL$ and modified $PL$ ones, simultaneous but reduced
errors are made: the characteristic surfaces are given with 20\%
overestimation (instead of $50\%$ with $PL$), while the viscous and thermal
increments are given with 9\% error (instead of 20\% with modified $PL$).

In general, the $JA$' and $PL$' models unsatisfactory account of parameters $%
\Sigma$ and $\Sigma^{\prime }$, will be at the origin of some errors in the
description of the propagation of transients at relatively short times or
short distances, whereas the modified $PL$' model description unsatisfactory
account of parameters $\alpha_0$ and $\alpha^{\prime }_0$, will be at the
origin of errors in the description of transients at longer times and
distances. To study this we compare in what follows different exact (fully
exact or asymptotically exact) and modelled time Green's functions for the
case of cylindrical circular tubes.

Let us first define the Green's functions and write the exact asymptotic
results that have been described in introduction and are more detailed in
Appendix.

\section{A simple definition of a Green's function; scaled form \label{SW}}

\subsection{Definition and scaled form}

A general method of defining and calculating a Green's function in an
infinite medium is by means of the effective frequency-dependent wavenumber $%
k$ in this medium. Let us define our Green's function $G(x,t)$ (or impulse
response) as a propagated Dirac delta impulsive signal $\delta (t)$ imposed
at $x=0$, or more precisely, as the inverse Fourier transform of the
propagation transfer function $G_{x}(\omega)=\exp (ikx)$:
\begin{equation}
G(x,t)=\int_{-\infty }^{\infty }\frac{d\omega }{2\pi }\exp \left[ -i\omega
t+ikx\right] .  \label{Green}
\end{equation}

Setting to zero the viscosity and thermal conduction coefficients, no
frequency dispersion arises. The wavenumber writes $k=\omega /c$ where $c$,
defined by the first Eq. (\ref{1a}), is the frozen speed of sound $c=c_{f}/%
\sqrt{\alpha _{\infty }}$. The Green's function (\ref{Green}) coincides with
a Dirac delta propagated at this velocity $c$: $G(x,t)=\delta (t-x/c)$.

Setting to nonzero values the viscosity and thermal conduction coefficients
the medium wavenumber $k$ writes,%
\begin{equation}
k=\frac{\omega }{c}\sqrt{\frac{\alpha (\omega )\beta (\omega )}{\alpha
_{\infty }}}=\frac{\omega }{c}\left[ 1+h(S_{T})\right] ,  \label{23}
\end{equation}%
making apparent a complex function $h(S_{T})$ that describes frequency
dispersion induced by the viscous and thermal relaxation processes. The
Green's function (\ref{Green}) writes,
\begin{equation}
G(x,t)=\int_{-\infty }^{\infty }\frac{d\omega }{2\pi }\exp \left[ -i\omega
(t-x/c)+i\frac{\omega }{c}h(S_{T})x\right] \text{ .}  \label{Green2}
\end{equation}

This Green's function $g_{x}(t-x/c)=G(x,t)$ now extends and flattens when
observed at positions $x$ more and more remote from the origin $x=0$.
Explicit asymptotic expressions of the Green's function defined in this
manner will now be obtained by considering high-frequency asymptotic
expansions of the wavenumber $k$ (meaning high-frequency asymptotic
expansions of the function $h$).

It will also be convenient to express these functions in scaled form, as
functions of a dimensionless delay time $\tau $ and a dimensionless position
variable $\xi $. To take but one example, before being Fourier-transformed,
the Green's function $G_{x}(\omega )=\exp (ikx)$ where $k$ is -- say --
given by the model $PL$, may be viewed as a function of:

(\textit{i}) one dimensionless frequency variable, e.g. $\Omega =\omega
\Theta $ where $\Theta $ is the characteristic viscous relaxation time given
by,
\begin{equation}
\Theta =\frac{\Lambda ^{2}\rho _{f}}{\eta },  \label{theta}
\end{equation}%
(hence $S_{T}^{2}=-i\Omega $),

(\textit{ii}) one dimensionless position variable, e.g.
\begin{equation}
\xi =\frac{x}{c\Theta },
\end{equation}%
(hence $\xi ^{-1/2}=\Lambda \sqrt{\rho _{f}c/(x\eta )}$ can be regarded as a
time domain Stokes number, when replacing $-i\omega $ by $c/x$ in Eq. (\ref%
{12})).

(\textit{iii}) a number of dimensionless parameters characteristic of the
form of the porous space but not of its absolute dimensions ($\alpha
_{\infty }$, $M$, $q$, $M^{\prime }$, $q^{\prime }$, and $\Lambda ^{\prime
}/\Lambda $), and,

(\textit{iv}) two dimensionless parameters characteristic of the fluid ($%
\gamma $ and $\Pr $).

Suppose that the parameters (\textit{iii}) and (\textit{iv}) of both the
medium and the fluid are held constant. Function $G_{x}(\omega )=\exp (ikx)$
then reduces to a function of $\Omega $ which is parameterized by $\xi $.
There follows that the shape of the corresponding time-domain function will
depend, in scaled form, on $\xi $ only, provided the time is counted in a
dimensionless manner, $e.g.$ for the time elapsed after the first arrival of
the signal,
\begin{equation}
\tau =(t-\frac{x}{c})/\Theta .  \label{tau}
\end{equation}%
But function $G(x,t)$ has the dimension of the inverse of time. Therefore,
\begin{equation}
\Theta G(x,t)=F_{s}(\xi ,\tau ),  \label{scaled}
\end{equation}%
with $F_{s}$ a scale-invariant function that depends on the form of the pore
space but not on its absolute dimensions:%
\begin{equation}
F_{s}(\xi ,\tau )=\int_{-\infty }^{\infty }\frac{d\Omega }{2\pi }\exp \left[
-i\Omega \tau +i\Omega \xi h(\sqrt{-i\Omega })\right] .  \label{K2}
\end{equation}%
Evidently, function $F_{s}$ will depend on the description of wavenumber or
function $h(S_{T})$: model $JA$, model $PL$, only some high-frequency terms
retained, or complete exact Zwikker and Kosten's.

\subsection{Asymptotic expansions \label{AE}}

Two successive closed-form analytical exact asymptotic expressions can be
derived for the frozen limit, using the series expansion of the wavenumber
for the case where the function $h(S_{T})$ is represented by its first, and
second order terms on $S_{T}^{-1}$. In the first case, using (Eqs. (\ref{hf}%
) and (\ref{23})), we write:
\begin{equation}
h=\frac{2n_{1}}{\sqrt{-i\Omega }}+o(1),
\end{equation}%
where,
\begin{equation}
2n_{1}=1+(\gamma -1)\frac{\Lambda }{\Lambda ^{\prime }\sqrt{\Pr }},
\end{equation}%
and obtain,
\begin{eqnarray}
\Theta G_{o(1)}(x,t) &=&0\text{ for }\tau <0,  \notag \\
\Theta G_{o(1)}(x,t) &=&F_{s1}(\xi ,\tau )\text{ for }\tau >0,
\label{maino1}
\end{eqnarray}%
where,
\begin{equation}
F_{s1}(\xi ,\tau )=\frac{1}{\sqrt{\pi }}\frac{n_{1}\xi }{\tau ^{3/2}}\exp %
\left[ -n_{1}^{2}\frac{\xi ^{2}}{\tau }\right] \text{ . }  \label{maino1'}
\end{equation}%
In the second case we write
\begin{equation}
h=\frac{2n_{1}}{\sqrt{-i\Omega }}+\frac{n_{2}}{-i\Omega }+o(2),
\end{equation}%
where,
\begin{equation}
2n_{2}=\frac{3\Lambda ^{2}}{\Sigma }-1+\frac{2(\gamma -1)\Lambda }{\Lambda
^{\prime }\sqrt{\Pr }}+\frac{(\gamma -1)\Lambda ^{2}}{\Lambda ^{\prime 2}\Pr
}\left( \frac{3\Lambda ^{\prime 2}}{\Sigma ^{\prime }}-3-\gamma \right) ,
\label{N2}
\end{equation}%
and obtain,
\begin{eqnarray}
\Theta G_{o(2)}(x,t) &=&0\text{ for }\tau <0,  \notag \\
\Theta G_{o(2)}(x,t) &=&F_{s1}(\xi ,\tau )\exp (-n_{2}\xi )\text{ for }\tau
>0.  \label{maino2}
\end{eqnarray}%
This yields\ an attenuation without distortion, corresponding to the
asymptotic results indicated in Eqs. (\ref{O2}--\ref{decay L}). This
analytical form of the Green's function, exhibiting a superimposed
exponential decay, is also a known result of musical acoustics \cite{polack}.

Concerning the relaxed limit, the purely diffusive Green's function $%
G_{diff} $ can be obtained by assimilating $\beta (\omega )$ with $\gamma $
and $\alpha (\omega )$ with its first leading term $\eta \phi /(-i\omega
\rho _{f}k_{0})$ in Eq. (\ref{bf}). In this relaxed d.c. approximation, the
wavenumber expands as:
\begin{equation}
k=k_{diff}=\frac{1}{c}\left( \frac{8\gamma }{M\Theta }\right) ^{1/2}\sqrt{%
i\omega }.
\end{equation}%
(we have used the relation (\ref{j15a})). The corresponding analytical form
of the Green's function is deduced from the known result given by Ref. \cite%
{landau}:
\begin{eqnarray}
\Theta G_{diff}(x,t) &=&0\text{ for }\tau ^{\prime }<0  \notag \\
\Theta G_{diff}(x,t) &=&F_{s,diff}(\xi ,\tau ^{\prime })\text{ for }\tau
^{\prime }>0,  \label{diff}
\end{eqnarray}%
where the notation $\tau ^{\prime }=t/\Theta $ is used to avoid confusion
with Eq. (\ref{tau}):
\begin{equation}
F_{s,diff}(\xi ,\tau ^{\prime })=\frac{1}{\sqrt{\pi }}\sqrt{\frac{2\gamma }{M%
}}\frac{\xi }{\tau ^{\prime 3/2}}\exp \left[ -\frac{2\gamma }{M}\frac{\xi
^{2}}{\tau ^{\prime }}\right] .  \label{diff'}
\end{equation}

\section{\protect\bigskip FFT computation of Green's functions
(full-frequency expressions)\label{FFTC}}

Green's functions $G(x,t)$ computed numerically by inverse FFT (fast Fourier
transform) of the transfer function $\exp (ikx)$, can always be obtained as
soon as exact or approximate full-frequency expressions of the wavenumber $k$
in Eq. (\ref{23}) are known. These FFT computations are to be carefully
done, however, as it is detailed below. In Appendix, the different results,
either analytical or numerical, corresponding to the asymptotic Green's
functions are summarized and compared with the exact results of the FFT.

For the case of arbitrary porous media, approximate models such as $JA$ or $%
PL$ may be used for $k(\omega )$, leading to the computation, by FFT, of
approximate Green's functions $G(x,t)$. For the case of simple workable
geometries (e.g., cylindrical circular aligned pores of identical radius)
exact expressions are available, leading to the computation, by FFT, of
exact Green's functions $G(x,t)$. For later use, it will be convenient to
distinguish and denote respectively by:

$G$ the exact (Zwikker and Kosten) Green's function;

$G_{JA}$ the full $JA$ Green's function ($q=q^{\prime }=1$);

$G_{PL}$ the full $PL$ Green's function ($q=q^{\prime }=3/4$);

$G_{mPL}$ the modified full $PL$ Green's function (defined by $PL$'s
expressions with $q=q^{\prime }=5/8$);

$G_{o(1)}$ the above asymptotic Green's function obtained by retaining in
the wavenumber the zero and first frozen terms;

$G_{o(2)}$ the above asymptotic Green's function obtained by retaining in
the wavenumber the zero, first and second frozen terms.

\subsection{The use of the asymptotic diffusive Green's function in the FFT
computation}

An important point is that the analytical diffusion solution $G_{diff}$
given by Eq. (\ref{diff}) is used in our FFT computations to improve the
accuracy of calculations. Indeed, the numerical computation of the
full-frequency exact and model Green's functions must be done by FFT with
some precautions: because of the importance of the higher frequencies on the
response shape, subtracting the diffusive (low frequencies) approximation
largely improves the results. Thus, instead of calculating directly the
inverse Fourier transform $G(x,t)$ of the function $\exp (ikx)$, we first
compute the inverse Fourier transform $G^{\prime }(x,t)$ of the function $%
\exp (ikx)-\exp (ik_{diff}x):$ The latter difference being zero at zero
frequency, and increasing smoothly then decreasing rapidly toward zero at
high frequency,  the function $G_{x}(\omega )$ is naturally windowed in the
frequency domain.  And then we use the relation $G(x,t)=G^{\prime
}(x,t)+G_{diff}(x,t)$, with $G_{diff}(x,t)$ as given by the Eqs. (\ref{diff}-%
\ref{diff'}) above (see Ref. \cite{kergo}). The validity of the FFT
computation has been checked for first order asymptotic expression $G_{o(1)}$%
, with an accuracy better than 1\%.

\subsection{Computed results: time responses \label{re}}

For the computation, the values of the parameters have been chosen close to
those of material M1 in Ref. \cite{fellah}. The porosity is $\phi =0.82$,
the flow resistivity $\sigma =196000Nm^{-4}s$, the permeability $k_{0}=\eta
/\sigma =9.225\;10^{-11}m^{2}$, corresponding to a radius $R=3$ $10^{-5}m$.
The temperature is $20%
%TCIMACRO{\U{b0}}%
%BeginExpansion
{{}^\circ}%
%EndExpansion
C$, and the Prandtl number \ $\Pr =0.71$. The characteristic viscous
relaxation time is $\Theta =6.10^{-5}s.$

\begin{figure}[h]
\centering \includegraphics[width=10cm]{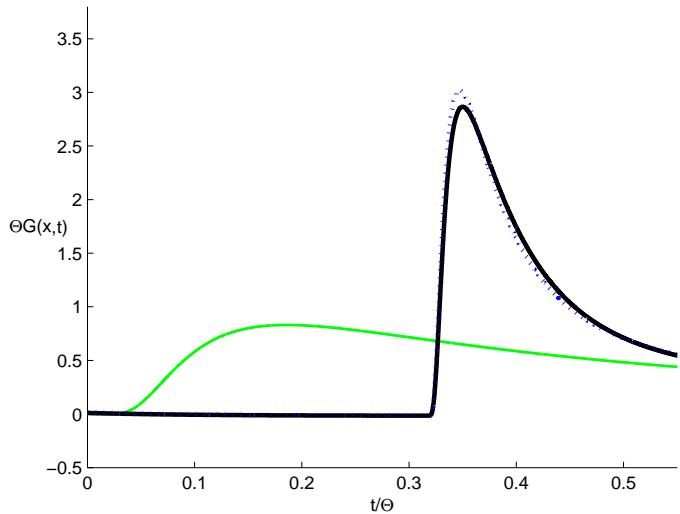}
\caption{{}Green's function $\Theta G(x,t)$ with respect to time $t/\Theta $
(both are dimensionless). The characteristic viscous relaxation time $\Theta
$ is defined by Eq. (\protect\ref{theta}). (Black) solid line: Zwikker and
Kosten formulae (Eqs. (\protect\ref{ZW1}) and (\protect\ref{ZW2})). (Blue)
dotted line: Pride-Lafarge description (Eqs. (\protect\ref{14}) to (\protect
\ref{15b})). (Green) thin solid line: diffusive limit (Eqs. (\protect\ref%
{diff}) and (\protect\ref{diff'})). For very long times, the diffusive limit
is reached.}
\label{fig1}
\end{figure}

The chosen length is $5.3$ $mm$ (the dimensionless length is $\xi =0.3152$).
Fig. \ref{fig1} shows the results for the dimensionless Green's function,
obtained by FFT. The Pride-Lafarge description is compared with the exact
Zwikker and Kosten formula. As expected, the full $PL$ description is very
satisfactory for short times (high frequencies) and long times (low
frequencies). For very long times, both descriptions reach the diffusive
(analytical) limit, $i.e.$ the Poiseuille behavior is reached.
\begin{figure}[h]
\centering \includegraphics[width=10cm]{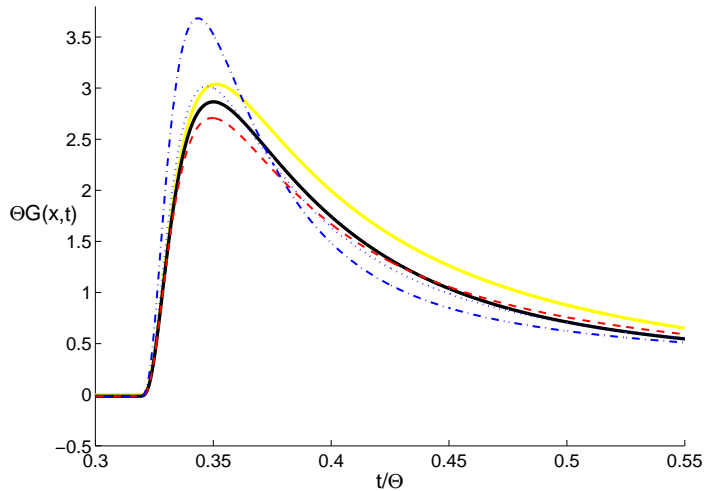}
\caption{{}Green's function $\Theta G(x,t)$ with respect to time $t/\Theta $
(zoom of figure \protect\ref{fig1}). (Black), solid line: Zwikker and Kosten
formula. (Blue), dotted line: $PL$ formula. (Red), dashed line: $PL$
modified formula. (Blue), mixed line: $JA$ formula. (Yellow), solid, pale
line: frozen $o(2)$'s Green's function $G_{o(2)}$ -- see Appendix, Eqs. (%
\protect\ref{maino1'}--$\,$\protect\ref{maino2}). }
\label{fig2}
\end{figure}

In order to emphasize these results, Fig. \ref{fig2} shows a zoom of the
previous figure, and other approximations have been added. The modified $PL$
Green's function, denoted $G_{mPL}$, can be compared to the $PL$ Green's
function. As expected, it is more accurate for short time (during the signal
rise) than the original $PL$ function, but it is less accurate for long
times, because the choice of the parameters $q$ and $q^{\prime }$ has been
done from the frozen limit instead of the relaxed limit. Otherwise the $JA$
model yields less accurate results than both $PL$ models. The asymptotic
Green's function $G_{o(2)}$ which uses the same number of parameters as $JA$
performs quite well. On Fig. 2 it is observed that the models $PL$ and $mPL$
are in error mainly in the region of the maximum; one elementary means to
improve the description of the bump shape of the response would be to take
the direct mean of the two models Green's functions. The response modelled
in this manner\footnote{%
Or, with almost undistinguishable results, with model $PL$ calculated with
values of $q$ and $q^{\prime }$ equal to the mean of their original and
modified $PL$ values.} would be very close to the exact one. Additional
results given later on (Fig.4) show however that this is a favourable
situation related to the value $\xi \simeq 0.3$ and that no very significant
improvement of model $PL$ is to be expected in this manner for other $\xi $
values.

\subsection{Computed results: Maximum values of the time response}

\begin{figure}[h]
\centering \includegraphics[width=10cm]{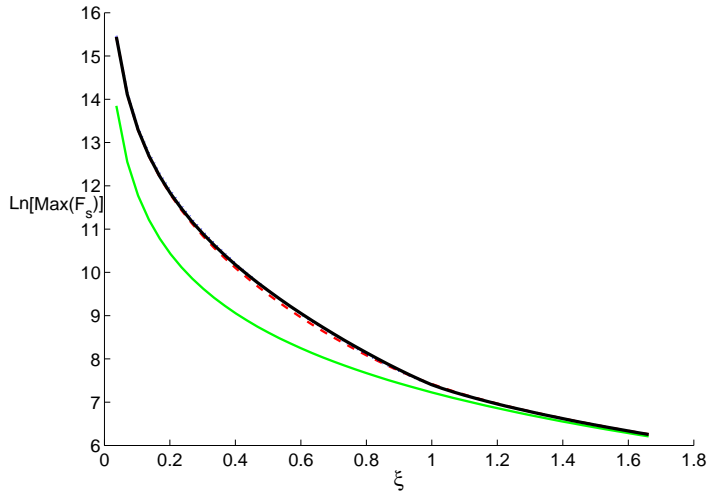}
\caption{{}Natural logarithm of the Green's function maximum. (Black) solid
thick line: exact Green's function $G$. (Red) dashed line: modified Green's
function $G_{mPL}.$ (Green) solid thin line: diffusive Green's function $%
G_{diff}$.(Blue) dotted line: $PL$ Green function $G_{PL}$ -- at this scale
this curve cannot be distinguished from the exact result, see Fig. \protect
\ref{fig4} for details.}
\label{fig3}
\end{figure}

The previous results are concerned with a fixed value of the parameter $\xi $%
, $i.e.$ a fixed value of the thickness of the material layer. In order to
compare the different descriptions for several values of $\xi $, we chose to
compare the maximum values of the time responses (there is a unique
maximum). For the asymptotic (frozen) expressions at the first orders, $%
F_{s1}$ and $F_{s2}$, the maximum values are given by an analytical
expression, obtained from Eq. (\ref{maino1'}):%
\begin{eqnarray}
Max(F_{s1}) &=&\frac{1}{\sqrt{\pi }}\left[ \frac{3}{2}\right]
^{3/2}e^{-(3/2)}\frac{1}{n_{1}^{2}\xi ^{2}}=\frac{0.2312}{n_{1}^{2}\xi ^{2}}%
\text{ ;}  \label{jkj} \\
Max(F_{s2}) &=&Max(F_{s1})\exp (-n_{2}\xi ).  \label{jkk}
\end{eqnarray}

\begin{figure}[h]
\centering \includegraphics[width=10cm]{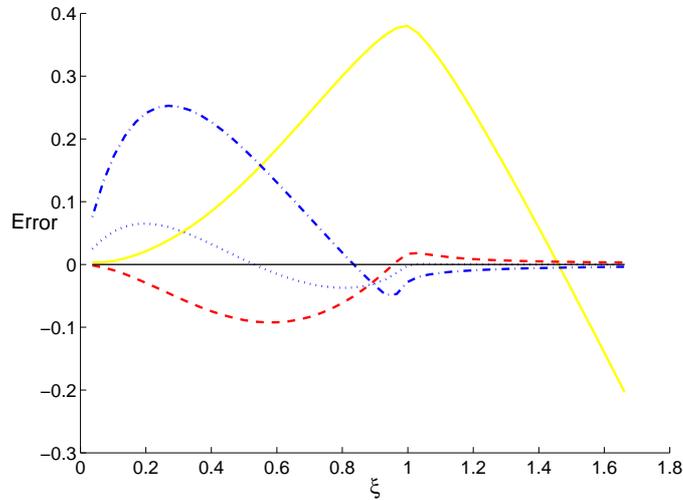}
\caption{{}Error on the natural logarithm of the Green's function maximum.
The different solutions are compared to the exact one, the error being the
difference between the corresponding value and the exact value. (Blue)
dotted line: $G_{PL}$. (Red) dashed line: modified $PL$ $G_{mPL}$. (Blue)
mixed line: $G_{JA}$. (Yellow) solid line: \ frozen $o(2)$'s Green's
function $G_{o(2)}$ -- see Appendix, Eqs. (\protect\ref{maino1'}$\,$--$\,$%
\protect\ref{maino2}). }
\label{fig4}
\end{figure}

For both orders the time $\tau _{\max }=2n_{1}^{2}\xi ^{2}/3$ of the maximum
is the same (this illustrates the attenuation without distortion effect
brought by the $O(2)$ terms). Fig. \ref{fig3} shows the result for the exact
Green's function $G$ and the two $PL$ models, $G_{PL}$ and $G_{mPL}$, as
well as the diffusive function $G_{diff}$. It shows the natural logarithm of
the maximum value of $F_{s}$ with respect to the dimensionless space
variable $\xi $. The two $PL$ models seem to be very good; however better
insight is found by subtracting the result corresponding to the exact
Green's function, as shown in Fig. \ref{fig4}. As expected, the $PL$
description is very good for long distances $\xi $, while the modified $PL$
description is better for small distances $\xi$. Otherwise both are better
than the $JA$ description. The transition range values of the distance $\xi $
is approximately between $0.036$ and $1.7$, corresponding to a range for the
time domain Stokes number $5.3\geq \xi ^{-1/2}\geq 0.75$. This range is
similar to that accepted for the frequency domain Stokes number defined by
Eq. (\ref{12}) (see e.g. Ref. \cite{keefe}).

%\clearpage

\section{Conclusion}

A simple analytic formula, Eq.(1), can be used to compute the $o(2)$
asymptotic Green's function in a rigid-framed porous medium. It differs from
the $o(1)$ asymptotic Green's function, by an exponential factor $\exp (-x/%
\mathcal{L})$ which describes an attenuation-without-distortion effect.

We have so far no rigorous statements concerning the geometrical surfaces $%
\Sigma$ and $\Sigma^{\prime }$ that determine the $O(2)$ terms in the
high-frequency limits (\ref{hf}) and then also determine the characteristic
decay length $\mathcal{L}$ Eq. (2). Nevertheless these parameters are known
for some geometries, such as cylindrical pores.

By specializing to this particular geometry, a contrasted situation has been
highlighted: while the models fail to give the parameters $\Sigma$, $%
\Sigma^{\prime }$, and thus, a correct description of the $o(2)$
attenuation-without-distortion effect, they are nevertheless capable to give
-- especially the model $PL$ -- a relatively precise description of the
complete Green's functions. Indeed, it is only at very short distances that
the asymptotic $o(2)$ Green's function is close to the complete Green's
function: its imperfect representation by the models is of no very
significant consequence.

In connection with this, we note that, when the normalized distance $\xi$
decreases, the maximum error of $PL$'s model occurs around $\xi=0.2$,
precisely when the $o(2)$ Green's function eventually starts to be valid
(see Fig. 4). This suggests that the small remaining errors of $PL$'s model
illustrated in Fig. 4 are mainly due to the misrepresentation of parameters $%
\Sigma$, $\Sigma^{\prime }$, and that the model would become almost exact if
modified to properly account for the latter parameters. The question of the
modification to be done remains open.

A problem of major interest is the use of the present investigation for the
inverse problem, $i.e.$ the determination of the parameters of a given
material. Regarding this, we have illustrated one simple fact: the
description of the time domain Green's functions is much more precise using
the full model expressions than using the asymptotic expressions, as often
done in practice.

This suggests that there is also an important potential of improvement of
the inverse methods of characterization based on recording transmitted and
reflected pulses on different thicknesses of a material, provided the full
expressions are exploited in the analysis -- we emphasize, in this respect,
the importance of the substraction of the diffusive solution when computing
the inverse FFT.

%\appendix
%\appendixpage
\renewcommand{\theequation}{A-\arabic{equation}}
% redefine the command that creates the equation no.
\setcounter{equation}{0} % reset counter
\setcounter{section}{0} % reset counter

\section*{APPENDIX: Asymptotic expressions (frozen limit)}

\section{Asymptotic $o(1)$'s and $o(2)$'s expressions of the Green$^{\prime
}s$ function}

Recall that, since we assume a smooth pore-surface interface, in the frozen
limit the product $\alpha (\omega )\beta (\omega )/\alpha _{\infty }$ in Eq.
(\ref{23}) expands in successive integral powers of $S_{T}^{-1}$ (see
footnote \ref{fract}). This has been done with dimensionless variables in
section \ref{AE}. The two coefficients $n_{1}$ and $n_{2}$ have been
obtained using the a priori expansions of the functions $\alpha (\omega )$
and $\alpha ^{\prime }(\omega )$ (Eqs. (\ref{hf}))$.$ For the case of
cylindrical circular pores, the result for $n_{2}$ has been given by Keefe%
\cite{keefe}, using the Zwikker and Kosten solution:%
\begin{equation}
\text{\ }n_{2}=1+\frac{\gamma -1}{\sqrt{\Pr }}-\frac{\gamma (\gamma -1)}{%
2\Pr },  \label{45}
\end{equation}%
(Eq (\ref{45}) also follows by putting the values (\ref{19}) in Eq. (\ref{N2}%
)). For the general case, $n_{2}$ requires the missing frozen $O(2)$
information $\Sigma $ and $\Sigma ^{\prime }$. It will not be given by the
asymptotic expansion, Eqs. (\ref{16}) and (\ref{18}), with either $JA$ or $PL
$ values of $q$ and $q^{\prime }$. As explained in section \ref{ffm}, in the
framework of the $PL$ model, it is not possible to have a good estimation of
the second order term, resulting in an expression for the coefficient $n_{2}$
which differs from Eq. (\ref{N2}):

\begin{equation}
2n_{2}=\frac{8(1-q)}{M}-1+\frac{2(\gamma -1)\Lambda }{\Lambda ^{\prime }%
\sqrt{\Pr }}+\frac{(\gamma -1)\Lambda ^{2}}{\Lambda ^{\prime 2}\Pr }\left(
\frac{8(1-q^{\prime })}{M^{\prime }}-3-\gamma \right) .  \label{42b}
\end{equation}
The case of cylindrical circular pores allows checking this. For this case,
the latter equation becomes:%
\begin{equation}
\text{ }n_{2}=\frac{1}{2}+\frac{\gamma -1}{\sqrt{\Pr }}-\frac{\gamma ^{2}-1}{%
2\Pr }.  \label{44}
\end{equation}%
This expression differs from Eq. (\ref{45}). For standard conditions in air,
the exact result for $n_{2}$ is $1.08$, while the approximated one is $0.29$%
, $i.e.$ more than three times smaller. Notice that $JA$ model would give a
negative estimate of $-1.27$ for it. These important discrepancies mean that
the models $JA$ and even $PL$ will not be able to describe the exact $o(2)$
attenuation effect.

\section{Asymptotic \textit{o}(2) wave equation\label{WE}}

An alternative to the closed-form Green's function obtained for this same $%
o(2)$ asymptotic frozen limit can be obtained using the 1D wave equation
that follows from Eqs. (1), $i.e.$ the following Helmholtz equation:%
\begin{equation}
\frac{d^{2}p}{dx^{2}}+\omega ^{2}\frac{\rho _{f}\alpha (\omega )\beta
(\omega )}{K_{f}}p=0.  \label{20}
\end{equation}%
%
%
%
%
%
%
%
%
%
%
%
%
%Using the $PL$ high-frequencies asymptotic limits (\ref{16}) and (\ref{18}) we get,
Using the high-frequencies asymptotic limits (\ref{hf}) we get,
\begin{eqnarray}
\alpha (\omega )\beta (\omega )/\alpha _{\infty } &=&1+\frac{2}{S_{T}}%
+(\gamma -1)\frac{2}{S_{T}^{\prime }}+\frac{3\Lambda ^{2}}{\Sigma S_{T}^{2}}+
\notag \\
&&(\gamma -1)\left( \frac{3\Lambda ^{\prime 2}}{\Sigma ^{\prime
}S_{T}^{\prime 2}}-\frac{4}{S_{T}^{\prime 2}}+\frac{4}{S_{T}S_{T}^{\prime }}%
\right) +o(2),  \label{exalim}
\end{eqnarray}%
%
%
%
%
%
%
%
%
%
%
%
%
%\begin{eqnarray}
%\alpha (\omega )\beta (\omega )/\alpha _{\infty } &=&1+\frac{2}{S_{T}}+(\gamma-1)\frac{2}{S_{T}^{\prime }}+\frac{8(1-q)}{MS_{T}^{2}}+  \notag \\ &&(\gamma -1)\left( \frac{8(1-q^{\prime })}{M^{\prime }S_{T}^{\prime 2}}-\frac{4%}{S_{T}^{\prime 2}}+\frac{4}{S_{T}S_{T}^{\prime }}\right) +o(2),\label{lim}
%\end{eqnarray}
%and this is the form of the exact asymptotic result, except of course for the $PL$ faulty values of $q$ and $q^{\prime }$ if we intend to use, for them, the Eqs. (\ref{15b}).

In the time domain, the corresponding asymptotic wave equation is written as
follows \cite{fellah}:%
\begin{equation}
\frac{\partial ^{2}p(x,t)}{\partial x^{2}}-\mathcal{A}\frac{\partial
^{2}p(x,t)}{\partial t^{2}}-\mathcal{B}\int_{0}^{t}\frac{\partial
^{2}p(x,t^{\prime })/\partial t^{\prime 2}}{\sqrt{t-t^{\prime }}}dt^{\prime
}-\mathcal{C}\frac{\partial p(x,t)}{\partial t}=0.  \label{we}
\end{equation}%
Comparison between (\ref{20}-\ref{exalim}) and (\ref{we}) shows that the
coefficients are given by:%
\begin{eqnarray}
\mathcal{A} &=&\frac{1}{c^{2}}=\frac{\rho _{f}\alpha _{\infty }}{K_{f}}\text{%
, \ \ }\mathcal{B}=4n_{1}\sqrt{\frac{1}{\pi }}\frac{1}{\Lambda c^{2}}\sqrt{%
\frac{\eta }{\rho _{f}}},  \label{1a} \\
\mathcal{C} &=&\frac{1}{c^{2}}\frac{\eta }{\rho _{f}}\frac{1}{\Lambda ^{2}}m%
\text{ \ with}  \label{1b} \\
m &=&\frac{3\Lambda ^{2}}{\Sigma }+(\gamma -1)\left[ \frac{4\Lambda }{%
\Lambda ^{\prime }\sqrt{\Pr }}+\left( -4+\frac{3\Lambda ^{\prime 2}}{\Sigma
^{\prime }}\right) \frac{\Lambda ^{2}}{\Lambda ^{\prime 2}\Pr }\right] .
\label{exaa1c}
\end{eqnarray}%
The relationship between the coefficients $m$ and $n_{1}$ and $n_{2}$ is:%
\begin{equation}
\text{ }m=2n_{2}+4n_{1}^{2}.  \label{43}
\end{equation}

Notice that by using $PL$ model one would arrive in the asymptotic
high-frequency limit to the same asymptotic wave Eq. (\ref{we}) but with the
following erroneous value of the index $m$:
\begin{equation}
m=\frac{8(1-q)}{M}+(\gamma -1)\left[ \frac{4\Lambda }{\Lambda ^{\prime }%
\sqrt{\Pr }}+\left( -4+\frac{8(1-q^{\prime })}{M^{\prime }}\right) \frac{%
\Lambda ^{2}}{\Lambda ^{\prime 2}\Pr }\right] .  \label{a1c}
\end{equation}
In particular, using $JA$ model, two important terms disappear as for this
case one sets $q=q^{\prime }=1$. Using this expression for the index $m$ in (%
\ref{1b}) corresponds to using the equations (14-16) of Fellah $et$ $al.$%
\cite{fellah}, who computed the Green's function for an infinite medium
described by the above wave equation (\ref{we}), by using the Laplace
transform method\footnote{\label{9f} Notice that in Ref. \cite{fellah}\ \
there was a mistake of a factor 2 in the term under the root in Eqs. (\ref%
{14}) and (\ref{14p}), without influence on further equations. Moreover\ the
last term $-4/S_{T}^{\prime 2}$ in Eq. (\ref{18}) was omitted, resulting in
a total coefficient of the term in $S_{T}^{\prime -2}$ in Eq (\ref{18})
equal to $+2$. Here, consistent with Eq. (\ref{18}), an additional term has
been included in the bracket in Eq. (\ref{a1c}).}.

\section{Comparison of the asymptotic expressions}

The FFT computations of the exact Green's function can be compared with the
following expressions:

$G_{o(1)}$ the (frozen) $o(1)$ Green's function;

$G_{o(2)}$ the (frozen) $o(2)$ Green's function (Eq. (\ref{maino2}) with $%
q=q^{\prime }=5/8$ in Eq. (\ref{42b}));

$G_{JAo(2)}$ the $JA$ $o(2)$ Green's function (Eq. (\ref{maino2}) with $%
q=q^{\prime }=1$ in Eq. (\ref{42b}));

$G_{PLo(2)}$ the $PL$ $o(2)$ Green's function (Eq. (\ref{maino2}) with $%
q=q^{\prime }=3/4$ in Eq. (\ref{42b})).

%With $q=q^{\prime }=5/8$, $q=q^{\prime }=1$, and $q=q^{\prime }=3/4$, the above expression (\ref{o2}) yields the Green's functions $G_{o(2)}(x,t)$, $% G_{JAo(2)}(x,t)$, and $G_{PLo(2)}(x,t)$, respectively.

Finally, the FFT computations can be compared to the solution of the
asymptotic $o(2)$ wave equation (\ref{we}), for both cases $q=q^{\prime
}=5/8 $ and $q=q^{\prime }=3/4$ (we again choose to compute these solutions
using FFT, with Eqs. (\ref{23}) and (\ref{exalim}), without expansion of Eq.
(\ref{23})):

$G_{WEo(2)}$ the (frozen) $o(2)$ Green's function ($q=q^{\prime }=5/8$);

$G_{WEPL}{}_{o(2)}$ the $PL$ $o(2)$ Green's function ($q=q^{\prime }=3/4$).

These solutions being $o(2)$ are expected to be very close to the
corresponding solutions obtained using the asymptotic $o(2)$ wavenumber.

\begin{figure}[h]
\centering \includegraphics[width=10cm]{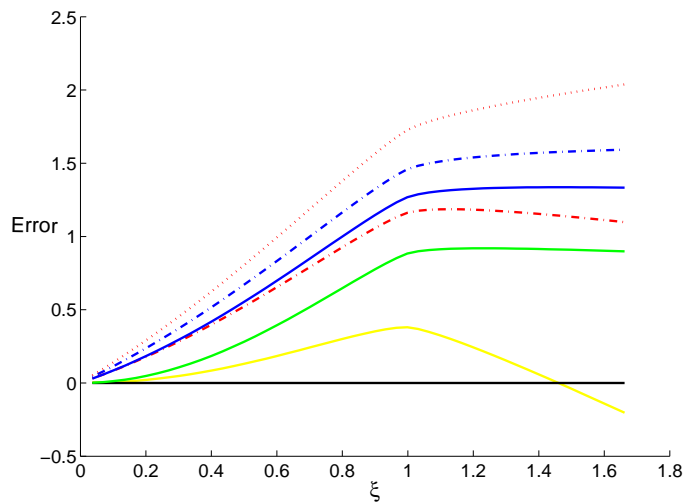}
\caption{{}Error on the natural logarithm of the Green's function maximum.
The different asymptotic solutions are compared to the exact one, the error
being the difference between the corresponding value and the exact value.
From the largest to the smaller error, the following curves represent
respectively: $G_{JAo(2)}$; $G_{o(1)}$; $G_{WEPLo(2)};$ $G_{PLo(2)}$; $%
G_{WEo(2)};$ $G_{o(2)}$, the latter being also shown in Fig. 4.}
\label{fig5}
\end{figure}

Results are plotted on Fig. \ref{fig5}. Notice that $G_{o(2)}$, the (frozen)
$o(2)$ Green's function, which is a very simple analytical expression, is
the best approximation and leads to interesting results, except at long
times.

The $PLo(2)$ description is slightly better than the first order of the
frozen asymptotic $G_{o(1)}$, but as expected, it is much comparable to the
latter, as it severely underestimate the decay length $\mathcal{L}$.
Otherwise, for short distances $\xi $, the second order of the asymptotic
wave equation solution $G_{WEo(2)}$ exhibits the expected convergence to the
results of the solution $G_{o(2)}$ based on wavenumber expansion at second
order. This convergence is lost for the comparable $PL$ model estimates $%
G_{WEPL}{}_{o(2)}$ and $G_{PLo(2)}$, as a result of using the faulty $PL$
coefficients ($q=q^{\prime }=3/4)$. For longer distances, the frozen
asymptotic $WE$ solution $G_{WEo(2)}$ appears to be less accurate than the
frozen asymptotic wavenumber solution $G_{o(2)}$: it is not easy to have an
interpretation for this result. The second order solution of the wave
equation, as presented in Ref. \cite{fellah} differs (by definition) by the
third order, with the solution based on wavenumber expansion at second
order, the latter being simpler to use in practice\footnote{\label{10f}
Looking at the calculation made in the frequency domain, the figure 3 of
this paper exhibits a ratio between the $PL$ description and the $JA$ one ($%
q=q^{\prime }=1$) that is almost independent of frequency: this is the
attenuation-without-distortion effect that is described by Eq. (\ref{maino2}%
) in our wavenumber-based asymptotic calculation.}.

\section*{Acknowledgments}

We wish to thank Bruno Lombard for fruitful discussions.

\end{document}